\documentclass{amsart}
\usepackage{amsmath, amsthm,  amsfonts, amscd, epsfig}

\newcommand{\tr}{{\text{tr}}}

\newcommand{\M}{\mathcal M}
\newcommand{\RR}{\mathcal R}
\newcommand{\Z}{\mathbb Z}
\newcommand{\R}{\mathbb R}
\newcommand{\C}{\mathbb C}

\font\frak=eufm10 at 11 pt
\newcommand{\su}{\mbox{\frak{su}}}

\theoremstyle{plain}

\newtheorem{theorem}{Theorem}[section]

\newtheorem{proposition}[theorem]{Proposition}

\theoremstyle{definition}

\theoremstyle{remark}

\begin{document}

\title[Monopoles]{Monopoles}
\author{Michael K. Murray}
\address
{Department of Pure Mathematics\\
University of Adelaide\\
Adelaide, SA 5005 \\
Australia}
\email{mmurray@maths.adelaide.edu.au}


\maketitle

\section{Introduction}
Monopoles are solutions of a first order partial differential
equation --- the Bogomolny equation.  They can be thought
of as approximated by static, magnetic particles in $\R^3$.  In these notes
we will consider
what monopoles are and some of the various approaches to
understanding them. In particular we will discuss
the spectral curve,  twistor theory,
  Nahm transform  and  rational map of a monopole.  A good starting reference
  for all this material is the book by Atiyah and Hitchin \cite{AtiHit} on
monopole scattering, in particular Chapter 16 `Background Material' from which
  these notes have borrowed notation.
See also the excellent survey by Sutcliffe \cite{Sut}.

\section{Monopoles in $\R^3$.}
We start with a Lie algebra  which we will assume  is
$\su(2)$, the  Lie algebra of all  skew-hermitian $2$ by $2$ matrices.
Let $A$ be a one-form with values
in $\su(2)$ so that $A = \sum_{i=1}^3 A_i  dx^i$ and each $A_i $
is a function $A_i \colon \R^3 \to\su(2)$.  The Higgs field $\Phi$ is a
function $\Phi \colon \R^3  \to\su(2)$.  The one-form $A$ can be thought of
as the connection one-form for a connection $\nabla = d + A$ on a
trivial $SU(2)$ bundle on $\R^3$.  The curvature of such a connection
is the two-form
$$
F_A = \frac{1}{2}\sum_{i , j =1}^3 F_{ij} dx^i \wedge dx^j
$$
where
$$
F_{ij} = [\nabla_i, \nabla_j] = \partial_i A_j - \partial_j A_i +[A_i, A_j].
$$
The connection $A$ can be used to covariantly differentiate the Higgs field
$\Phi$ to obtain
$$
\nabla_A \Phi = \sum_{i=1}^3 (\partial_i\Phi + [A_i, \Phi]) dx^i.
$$

A monopole is a connection $A$ and a Higgs field $\Phi$ satisfying the
Bogomolny equations and some particular boundary conditions.
The Bogomolny equations are:
\begin{equation}
	\label{eq:bog}
	F_A = * \nabla_A\Phi
\end{equation}
where $*$ is the Hodge star or duality operator from one-forms to
two-forms defined by
\begin{align*}
	*dx^1 &= dx^2 \wedge dx^3\\
	*dx^2 &= dx^3 \wedge dx^1 \\
	*dx^3 &= dx^1 \wedge dx^2.
\end{align*}

If $A$ and $B$ are elements of $\su(2)$ let $\langle A, B \rangle$  denote the
invariant form $\langle A, B \rangle = -\tr(AB^t)$. Then the energy 
density of a
pair  $(A, \Phi)$ is defined by
\begin{equation}
	\label{eq:energy}
	e(A, \Phi) = \frac{1}{2}|F_A|^2 + \frac{1}{2}|\nabla_A\Phi|^2
	\end{equation}
where
$$
|F_A|^2 = \sum_{i<j} \langle F_{ij} ,  F_{ij} \rangle
$$
and
$$
|\nabla_A\Phi|^2 =\frac{1}{2}\sum_{i}\langle\nabla_i\Phi,\nabla_i\Phi\rangle.
$$

The Yang-Mills-Higgs action of a pair $(A, \Phi)$ is the
integral over three space of the energy density:
\begin{equation}
	\label{eq:action}
{\mathcal L}(A, \Phi) = \int_{\R^3} e(A, \Phi) d^3x.
\end{equation}
If $B_R$ is a ball of radius $R$ integrating by parts shows that
$$
\int_{B_R} e(A, \Phi) d^3x  =\int_{B_R} \frac{1}{2} |F_A \pm *\nabla_A\Phi|^2
  d^3x  \mp \int_{S^2_R} \langle F_A , \Phi \rangle
  $$
where $S^2_R$ is the sphere of radius $R$.  If the limits of
all these integrals exist as $R \to \infty$ we obtain
$$
{\mathcal L}(A, \Phi)   =\int_{\R^3}  \frac{1}{2} |F_A \pm *\nabla_A\Phi|^2
  d^3x \mp \lim_{R \to \infty} \int_{S^2_R} \langle F_A , \Phi\rangle .
  $$
From this we easily deduce that the minima of the Yang-Mills-Higgs functional
are  solutions of the Bogomolny equations \eqref{eq:bog} or the 
anti-Bogomolny equations  $F_A = - * \nabla_A\Phi$.
As changing $\Phi$ to $-\Phi$ changes a solution of the Bogomolny 
equations to a solution of the anti-Bogomolny equations  we 
concentrate our attention on the former. 

The Bogomolny equations are invariant under {\em gauge transformations}
that is replacing $(A, \Phi)$ by
$$
(gAg^{-1} + gd(g^{-1}), g\Phi g^{-1})
$$
where $g \colon\R^3 \to SU(2)$.  The energy density \eqref{eq:energy} 
is also invariant
under gauge transformations.  When we talk about a monopole we are really
talking about an equivalence class of $(A, \Phi) $ under gauge
transformations.

The boundary conditions imposed on a  monopole are primarily that
the energy density \eqref{eq:energy} should have  finite integral --- 
that is the action
\eqref{eq:action} is finite.
There are some additional
technical conditions that we will not be concerned with.  It is
believed, in any case,  that these can all be deduced from finiteness 
of the action
and the Bogomolny equations.  From these boundary conditions we can
deduce that, after a suitable gauge transformation, we can arrange for
the Higgs field to have a limiting value at infinity
$$
\Phi^\infty(u) = \lim_{t \to \infty} \Phi(tu)
$$
where $u \in S^2$.  The boundary conditions can be used to show that
the eigenvalues of the Higgs field at infinity are independent of the
direction $u \in S^2$.  In the case of $SU(2)$ this is equivalent to
the fact that $|\Phi(u)|^2 =$ constant for all $u$.

It is easy to show that if $c > 0$ and $(A, \Phi)$ solves the
Bogomolny equations then
$$
(\hat A(x), \hat\Phi(x)) = (cA(x/c), c\Phi(x/c))
$$
also solves the Bogomolny equations. So we may as well
normalise the Higgs field so that $|\Phi(u)|^2 =1$ for all directions
$u$. Because the Lie algebra $\su(2)$ is three dimensional
the Higgs field at infinity is   a map
$$
\Phi^\infty \colon S^2 \to S^2  \subset\su(2).
$$
The space of all continuous maps $S^2 \to S^2$ breaks up into
connected components labelled by a winding number $k \in \Z$ just
as for maps $S^1 \to S^1$.

Because of this boundary condition we can arrange by a gauge
transformation for any Higgs field to satisfy
$$
\lim_{t \to \infty} \Phi(0, 0, t) = \frac{1}{\sqrt{2}}
\left(\begin{matrix} i & 0 \\ 0 & -i
\end{matrix}\right).
$$
We call such a Higgs field {\em framed}.
We define the moduli space of all monopoles of charge $k$,  $\M_k$, to
be  the space of all $(A, \Phi)$ solving
the Bogomolny equations, satisfying the appropriate boundary
conditions, with the Higgs field framed and modulo the action of
gauge transformations satisfying
\begin{equation}
	\label{eq:framedgauge}
\lim_{t \to \infty} g(0, 0, t) = \left(\begin{matrix} 1 & 0 \\ 0 & 1
\end{matrix}\right).
\end{equation}
If $k \leq 0$ then
$\M_k = \emptyset$, otherwise  $\M_k$
is a manifold of dimension $4k$.

Notice that the Bogomolny equations are translation invariant.
Moreover, because of the way we have defined the framing,
the group of all diagonal matrices (a copy of the
circle group $S^1$) acts by constant gauge transformations on
$\M_k$. Hence
$\R^3 \times S^1$ acts on $\M_k$.

If $k =1$ there is, up to this  action of $\R^3 \times S^1$,
a unique monopole, the Bogomolny,
Prasad, Sommerfield (BPS) monopole, given by
\begin{align*}
\Phi(x) &= \left(\frac{1}{r} - \frac{1}{\tanh r}\right) \frac{e}{r}\\
  A(x) &= \left(\frac{1}{\sinh r} - \frac{1}{r} \right)\frac{[e,de]}{r}
  \end{align*}
where $r = |x|$ and $e(x) = \sum_{i=1}^3 x^i e^i$ for an orthonormal
basis  $e^1$, $e^2$, $e^3$  of $\su(2)$.

\begin{figure}\label{fig:energy}
	\begin{center}
	\includegraphics[scale=0.3]{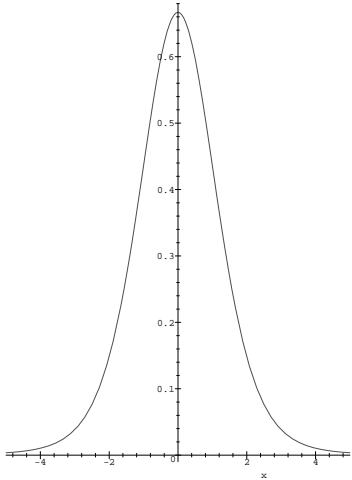}
\end{center}
\caption{Energy density for $k=1$, BPS monopole.}
\end{figure}

  The other $k=1$ monopoles are obtained by acting by
$\R^3 \times S^1$  so that $\M_1 = \R^3 \times S^1$.  It is easy to
calculate the energy density of the BPS monopole as follows.  For any monopole
there is a useful formula
$$
e(A, \Phi) = \sum_{i=1}^3 \frac{\partial^2\phantom{x^i}} {(\partial
x^i)^2} \langle \Phi, \Phi \rangle
$$
which can be proved using the Bogomolny equations and the Bianchi
identity.  From this we find that for the BPS monopole:
$$
e(A, \Phi) = \frac{6}{\tanh^4 r} -\frac{8}{\tanh^2 r} + 2 + \frac{2}{r^4} -
\frac{8}{r\tanh^3 r} + \frac{8}{r\tanh r}
$$
where $r = |x|$. Clearly the energy density is spherically symmetric 
about the origin in $\R^3$. If we plot it along the $x$-axis we obtain
Figure \ref{fig:energy}
and we see that the energy is concentrated around the origin. We think
of the monopole as a particle located at the origin.

The BPS monopole was discovered by Prasad and Sommerfield in 1975
\cite{PraSom}.  For some time this was the only monopole known and it
was unclear whether higher charge monopoles existed.  In 1977 Manton
\cite{Man} showed that to first order the forces between two
monopoles, due the Higgs field and the connection, cancelled. This
lead to the conjecture that stable higher charge monopoles would exist.
Weinberg in 1979 \cite{Wei} calculated that the dimension of the moduli
space of monopoles would be $4k$ if it was non-empty.
  Finally in 1981 Taubes \cite{JafTau} showed that
the moduli space was non-empty.

It is useful to briefly discuss Taubes' proof as it reveals some of
the interesting structure of the moduli space.
Taubes started with $k$ points a long way apart in $\R^3$ surrounded
by  very large balls.  At the centre of each of these balls he placed a BPS
monopole  and joined these all together
smoothly.  The result is not, of course, a solution to the Bogomolny
equations but it is close to one.   Such an approximate
solution can be used as the initial stage of an iterative scheme which
converges to a real solution.

The solutions produced  by Taubes are on the
`edge' of the moduli space.  That is,  the moduli space $\M_k$ has a
compact region outside of which   the monopoles
approximate a superposition BPS monopoles at $k$ points.  In 
particular the energy
density will be concentrated at these $k$ points.   One might also
expect that the Higgs field has a zero at each of these $k$ points but
this has not been proved.  Note that thinking of the charge $k$
monopoles as $k$ BPS monopoles gives an explanation of the dimension
of the moduli space: each constituent BPS monopole contributes its
position (three parameters) and a phase, which is a point on the circle,
making a total of four parameters.

It is important to note that when the charge is greater than one we
cannot associate to every monopole $(A, \Phi)$ a collection
of $k$ points which are the locations of the $k$ particles
we think of as its constituents. We can however associate sensibly
to a $k$ monopole a {\em centre of mass} or location \cite{HitManMur} which we
define below.

The analysis of monopoles directly in terms of the
connection and Higgs field on $\R^3$, for example the definition of
the location of a monopole, while possible, is difficult.  Part of this
difficulty stems from the  infinite dimensional symmetry group of 
gauge transformations.
Research on monopoles has focused on various transformations
which are designed to construct some other mathematical data equivalent to the
monopole. Study of these data  then, hopefully, sheds light on the original
monopole.  This process is particularly useful if the object produced
is an invariant of the monopole, something which does not change under
gauge transformation.    The transforms we will discuss below are:
\begin{itemize}
\item the spectral curve which is an invariant of the
monopole,
\item Hitchin's adaption of the twistor transform
which relates monopoles to holomorphic bundles on mini-twistor
space,
\item the Nahm transform which relates monopoles to solutions of
an ordinary differential equation on the interval $(-1, 1)$, and
\item the  rational map which is another invariant of the monopole.
\end{itemize}

\section{The spectral curve}
\label{sec:spectral}

Let $\gamma(t)$ be an oriented line in $\R^3$. This can be
  put in the form
$$
\gamma(t) = v + tu
$$
with vectors $u$ and $v$ determined uniquely by the requirement
that $|u|=1$, $\langle u, v \rangle = 0$ and $u$ points in the
direction of the orientation.  Along the line $\gamma$ we have
Hitchin's differential equation \cite{Hit}
\begin{equation}
	\label{eq:Hitchin}
  \left(\frac{d\phantom{t}}{dt} +
  \sum_{i=1}^3 u^iA_i(\gamma(t))
  -i\Phi(\gamma(t)) \right) s(t) = 0.
\end{equation}
This is an ordinary differential equation so it has a two-dimensional
space of solutions $E_\gamma$ depending, of course, on $\gamma$.

Notice that by a gauge transformation we can
arrange, for any given line $\gamma$, that $\sum_{i=1}^3 u^i A_i = 0$
so we can essentially disregard this term.
The boundary conditions can be used to show that we can expand the
Higgs field along any line as
\begin{equation*}
	\label{eq:Higgs}
	\Phi(\gamma(t)) = \left(\begin{matrix} i & 0 \\ 0 & -i \end{matrix}
\right)  - \frac{1}{2t}\left(\begin{matrix} ik & 0 \\ 0 & -ik \end{matrix}
\right) + O\left(\frac{1}{t^2}\right)
\end{equation*}
where the $k$ here is the monopole charge.  We want to consider the
Hitchin equation as a perturbation of a modified Hitchin equation:
\begin{equation}
	\label{eq:modHitchin}
  \left(\frac{d\phantom{t}}{dt}
+\left(\begin{matrix} 1 & 0 \\ 0 & -1 \end{matrix}
\right)  - \frac{1}{2t}\left(\begin{matrix} k & 0 \\ 0 & -k \end{matrix}
\right) \right) s = 0.
\end{equation}
The modified equation is the Hitchin equation with the
$o(1/t^2)$ term in the Higgs field expansion set to zero.
We use this to study the behaviour of solutions of the Hitchin
equation.   The solutions of \eqref{eq:modHitchin} are given by
\begin{gather}
s_1(t) = \left(\begin{matrix} t^{k/2}e^{-t} \\ 0 \end{matrix}\right)\\
s_2(t) = \left(\begin{matrix} 0 \\ t^{-k/2}e^{t} \end{matrix}\right).
\end{gather}
Asymptotic analysis of
ordinary differential equations shows that for any line there
  are solutions $s_1(t)$
and $s_2(t)$ of the Hitchin equation \eqref{eq:Hitchin} which behave
asymptotically like the solutions to the modified Hitchin equation, 
that is they satisfy
\begin{align*}
	\lim_{t \to \infty}  t^{-k/2}e^t s_1(t) &=
\left(\begin{matrix} 1 \\ 0 \end{matrix}\right)\\
\lim_{t \to \infty}  t^{k/2}e^{-t} s_2(t)   &=
\left(\begin{matrix} 0 \\ 1 \end{matrix}\right).
\end{align*}
Similarly there are solutions that decay and blow up exponentially
as $t \to -\infty$. For the modified Hitchin equation a solution that
blows up (decays) at one end of the line decays (blows up) at the
other end. In general this will not be true of solutions of the
Hitchin equation. In particular asymptotic analysis tells us that 
there will be a
ball in $\R^3$ of radius $R > 0$ with the property that if a line
lies outside the ball then the solutions $s_1(t)$ and $s_2(t)$
behave like the solutions to the modified Hitchin equation, that is
$s_1(t)$ decays as $t \to -\infty$ and $s_2(t)$ blows up as $t \to
-\infty$.

We expect then that lines which {\em do not} exhibit this behaviour
are somehow close to the monopole.  We call a line $\gamma$ a {\em
spectral line} if there is a solution to the Hitchin equation which
decays at both ends.  We call the set of all spectral lines the
{\em spectral curve} of the monopole.   It is easy to see  that
being a spectral line for a monopole is independent of gauge
transformations so the spectral curve is an invariant of the monopole.

It is not difficult to show that for the BPS monopole located at the
point $x \in \R^3$ the spectral lines are exactly the lines passing
through $x$. Note that this is a two-dimensional set, indeed a copy
of $S^2$.  This is more generally true: the  spectral
curve is always a two-dimensional family of lines.  To say more about 
the structure
of the spectral curve we need to consider the set of all oriented
lines in $\R^3$.

The importance of the spectral curve is the following theorem of
Hitchin:
\begin{theorem}[Hitchin \cite{Hit}]
	\label{th:spectral}
	If monopoles $(A, \Phi)$ and $(A', \Phi')$ have
	spectral curves $S$ and $S'$ and $S = S'$ then $(A, \Phi)$ is an
	unframed gauge
	transform of $(A', \Phi')$.
\end{theorem}
Note that  $S=S'$   means
equality of the corresponding sets of oriented lines in $\R^3$ and that
an unframed gauge transformation is one without
the requirement that $\lim_{t \to \infty} g(0, 0, t) = 1$ (c.f.
equation
\eqref{eq:framedgauge}). Theorem \ref{th:spectral} tells us that
the  spectral curve determines the monopole up to the
action of the circle on the moduli space. It is possible to also add
a small amount of additional data to the spectral curve which captures this
phase, we shall not concern ourselves with that here.

\section{The twistor theory of monopoles}
As we have discussed above each oriented line in $\R^3$ is determined
uniquely by vectors $u$ and $v$ satisfying $|u|=1$, $\langle u, v \rangle = 0$.
It follows that the set of all oriented lines is the tangent bundle
to the two-sphere:
$$
TS^2 = \{ (u, v) \mid |u|=1, \langle u, v \rangle = 0 \}.
$$
This is often called the {\em mini-twistor} space of $\R^3$.
Mini-twistor space is naturally a  complex manifold and
we can introduce co-ordinates
$(\eta, \zeta)$ on the open subset where $u \neq (0,0,1)$ by letting
\begin{equation}
    \label{eq:etazeta}
    \zeta = \frac{u^1 + i u^2}{1-u^3}  \quad\text{and} \quad
 \eta = (v^1 + i v^2) + 2v^3 \zeta + (-v^1 + iv^2)\zeta^2.
\end{equation}
The relationship between mini-twistor space and $\R^3$ is
summarised by the equation:
\begin{equation}
	\label{eq:minitwistor}
    \eta = (x^1 + ix^2) + 2 x^3 \zeta + (-x^1 + ix^2) \zeta^2.
\end{equation}

If we hold $(\eta, \zeta)$ fixed then the $(x^1, x^2, x^3)$ satisfying
\eqref{eq:minitwistor} define a line in $\R^3$. On the other hand if we
hold $(x^1, x^2, x^3)$ fixed then the $(\eta, \zeta)$ satisfying
\eqref{eq:minitwistor} parametrise the set of all lines through the
point $x=(x^1, x^2, x^3)$ which is a copy of $S^2$ inside $TS^2$.

Mini-twistor space has an
involution $\tau \colon TS^2 \to TS^2$ which sends each oriented
line to the same line with opposite orientation. In co-ordinates 
this is given by:
$$
\tau(\eta, \zeta) = \left(-\frac{\bar\eta}{\bar\zeta^2}, - 
\frac{1}{\bar\zeta} \right).
$$
Because $\tau$  is like a  conjugation it is called the   {\em real
structure}.  The set of all lines through the point $x$ is real in the
sense that it is fixed by the real structure.

We can now state the basic result concerning the spectral curve. It is
a subset of $TS^2$ defined by an equation of the form
\begin{equation}
	\label{eq:spectralcurve}
p(\eta, \zeta) = \eta^k + a_1(\zeta)\eta^{k-1} + \dots + a_k(\zeta) = 0
\end{equation}
where each of the $a_i(\zeta)$ is a polynomial of degree $2i$.

Note that by no means every such curve is the spectral curve of a
  monopole. One constraint is immediate from our definition. If a line
  is a spectral line then so also is the line with the opposite
  orientation.  So the spectral curve is {\em real}, that is, fixed
  by the real structure. But more is true.
  The family of real curves defined by equations of the form
  \eqref{eq:spectralcurve} is $(k-1)^2 -1 $ real dimensional whereas
  the moduli space of monopoles is $4k$ dimensional.  So there have to
  be further
  constraints on the spectral curve. In particular a certain holomorphic
  line bundle must be trivial when restricted to the spectral curve. It
  is possible to say quite precisely what the other constraints are and
hence to say, in principle, which spectral curves give rise to
monopoles \cite{Hit1}.

Spectral curves can be used to deduce a number of useful facts about
monopoles.  It is easy to show that the only real curves of the type
\eqref{eq:spectralcurve} for $k=1$ are those of the form
\eqref{eq:minitwistor} for some point $(x^1, x^2, x^3)$. Hence the BPS
monopoles are the only charge one monopoles.

The
coefficient $a_1(\zeta)$ in \eqref{eq:spectralcurve} defines a real
curve and hence has the form
$$
a_1(\zeta) =    (x^1 + ix^2) + 2 x^3 \zeta + (-x^1 + ix^2) \zeta^2
$$
for some point $(x^1, x^2, x^3)$.
in $\R^3$. This point is called the centre of the
monopole.   If $(u, v_1), \dots, (u, v_k)$ is a
collection of $k$ parallel lines let us define their average to be the
line $(u, (1/k)\sum_{i=1}^k v_i)$.  Notice from definition of $\eta$
\eqref{eq:etazeta}
that if these lines have complex co-ordinates $(\eta_1, \zeta), \dots,
(\eta_k, \zeta)$ then their average has complex co-ordinates
$((1/k)\sum_{i=1}^k \eta_i, \zeta)$. If we fix a particular direction 
in $\R^3$, that
is fix a $\zeta$ and look for all the spectral lines in that
direction we are finding all the $\eta$ satisfying  a degree $k$ polynomial
and hence there are generically $k$ of them.  If we take the average of
all these lines then it will pass through the monopole centre.  This
gives us a way of defining the centre entirely in $\R^3$. Take the
average of the spectral lines in each direction in $\R^3$, the
resulting family of lines will (nearly) all intersect in a single point,
that point is the centre of the monopole.

The definition of the spectral curve of a monopole clearly preserves
the action of the rotations and translations of $\R^3$ and this
gives us a way of looking for monopoles with particular symmetries.
We look first for spectral curves with these symmetries. This approach
can be used to show that the only spherically symmetric monopole is
the  BPS monopole at the origin. It was  used by Hitchin \cite{Hit1}
to classify the axially symmetric monopoles and more recently
\cite{HitManMur} to
find monopoles with symmetry groups those of the regular solids.

The various properties of the spectral curve such as the form
of equation \eqref{eq:spectralcurve} are proved by using Hitchin's
twistor transform.  Hitchin \cite{Hit} introduces the vector space $E_\gamma$
of all solutions to his equation \eqref{eq:Hitchin}. This is a
two-dimensional space and the collection of them all
defines a complex vector bundle $E$ over the mini-twistor space $TS^2$.
Hitchin shows  that the Bogomolny equations
imply that $E$ is a holomorphic vector bundle and moreover the
monopole can be recovered from knowing $E$.  The boundary conditions
of the monopole then enter by noting that there are two distinguished
holomorphic sub-bundles $E^+$ and $E^-$ of solutions to Hitchin's
equation which decay at $+$ and $-$ infinity.   The spectral curve is
the set where these line bundles coincide.  Algebraic geometry can
then be used to prove Theorem \ref{th:spectral} and that the
spectral curve satisfies an equation of the form
\eqref{eq:spectralcurve}.
Various constraints on the spectral curve also follow from the
twistor theory.  The precise constraints that a
curve must satisfy to be the spectral curve are given in \cite{Hit1}.
The proof that a spectral curve satisfying these constraints comes
from a monopole  requires the Nahm transform which we consider next.

\section{The Nahm transform and Nahm's equations}
An alternative point of view on monopoles comes via Nahm's
adaption of the Atiyah, Drinfeld, Hitchin, Manin construction
of instantons \cite{Nah}.  Nahm considers a Dirac operator on $\R^3$ coupled to
the monopole. In more detail let $\sigma^i$ be an orthonormal
basis for the Lie algebra $\su(2)$.  This particular $\su(2)$ should be
regarded as different to the monopole $\su(2)$ in which the connection
and Higgs field take values.  It is, in fact, the Lie algebra of the spin group of the
group of rotations of $\R^3$.  The  Dirac operator $D_z$ is defined by
$$
D_z = \sum_{i=1}^3 \sigma^i \nabla_i - (\Phi + iz)
$$
and acts on  $\C^2 \otimes \C^2$ valued functions on $\R^3$. The 
first $\C^2$ is the
spin space on which the $\sigma^i$ act and the second is the space on
which the $A_i$ and $\Phi$ act.  Here $z$ is any real number.
We also define an adjoint
$$
D^*_z = \sum_{i=1}^3 \sigma^i \nabla_i + (\Phi + iz).
$$
If we compute the composition the Bogomolny equations
show us that
$$
D_z D^*_z = \sum_{i=1}^3 \nabla_i\nabla_i - (\Phi+iz)(\Phi + iz)
$$
which is a positive operator and hence has no $L^2$ kernel.
 From this we conclude that $D^*_z$ has no $L^2$ kernel. An $L^2$
index theorem of Callias shows that $D_z$ has index $k$ if
$-1 < z < 1$ and $0$ otherwise. Hence it follows that
$D_z$ has a $k$ dimensional $L^2$ kernel $N_z$ if $-1 < z < 1$. The point
of view we wish to adopt  is that $N_z$ is a $k$ dimensional vector bundle
over the interval $(-1, 1)$.

We are interested in sections of this vector bundle, that is functions
$$
\psi \colon (-1, 1) \times \R^3 \to \C^2 \otimes \C^2.
$$
which satisfy  $D_z \psi(z, x)=0$ for every $z \in (-1,1)$.
Choose $k$ of these $\psi^1, \dots, \psi^k$ so that for each $z$ they
span $N_z$. Moreover choose them so that they are orthonormal
$$
\int_{\R^3} \left(\psi^i, \psi^j\right) d^3x = \delta^{ij}
$$
and satisfy
$$
\int_{\R^3} \left(\psi^i , \frac{\partial\psi^j}{\partial z}\right) 
d^3x = 0
$$
for all $i,j= 1 \dots, k$. Here, of course, 
$\delta^{ij}$ is the Kronecker delta which is zero unless $i=j$ when 
it 
is one.    Notice that there is no obstruction to satisfying these extra
conditions.  We can use
Gramm-Schmidt orthogonalisation for the first and the second is just
solving  an ordinary differential equation on $(-1, 1)$.

Now we define three $k$ by $k$ matrix functions of $z$ by
$$
T_a^{ij}(z) = \int_{\R^3} \left(\psi^i, x^a\psi^j\right) d^3x
$$
for $i,j = 1 \dots, k$ and $a=1,2,3$. 
The remarkable thing about this Nahm transformation is that
these matrix valued functions  satisfy some simple ordinary
differential equations, called  Nahm's equations
\begin{align*}
\frac{dT_1}{dz} &= [T_2, T_3]\\
\frac{dT_2}{dz} &= [T_3, T_1]\\
\frac{dT_3}{dz} &= [T_1, T_2].
\end{align*}

It is possible to cast Nahm's equations into Lax form and solve
them by the Krichever method. We will consider how to do this as it
connects us again with spectral curves.

Define
$$
A(\zeta) = (T_1 + iT_2) + 2T_3 \zeta + (-T_1 + iT_2) \zeta^2
$$
and $A_+(\zeta) =i T_3 - (iT_1 + T_2)\zeta$. Then Nahm's equations
are equivalent to
$$
\frac{dA}{dz} = [A_+, A]
$$
which is in Lax form.  Now consider the curve $S_z$ in $\C \times \C$
defined by
\begin{equation}
    \label{eq:nahmcurve}
    \det(\eta - A(\zeta)) = 0.
\end{equation}
We have
\begin{proposition}
	The curve $S_z$ is independent of $z$.
	\end{proposition}
\begin{proof}
	Choose an eigenvector $v$ of $A(\zeta)$ with eigenvalue $\lambda$ at
	some $z=z_0$. Now evolve $v$ with the differential equation
	$$
	\frac{dv}{dz} = A_+ v
	$$
	and consider
\begin{align*}
	\frac{d(Av)}{dz} &= A_+Av - AA_+v + AA_+v \\
	&= A_+(Av).
	\end{align*}
	It follows that
	$$
	\frac{d\phantom{z}}{dz}\left(Av - \lambda v \right) = 0
	$$
	and because $Av - \lambda v$ vanishes at $z_0$ it must
	vanish everywhere.   We conclude that the eigenvalues
	of $A(\zeta)$ are independent of $z$ and hence that $S_z$ is
	independent of $z$.
	\end{proof}

If we identify the $(\eta, \zeta) $ in \eqref{eq:nahmcurve} with  the 
co-ordinates \eqref{eq:etazeta} on mini-twistor space we realise $S_z$ as a curve in 
mini-twistor space.  This is, in fact, a natural thing to do.  
It is a remarkable fact \cite{HitMur} that the curve $S = S_z$
defined via Nahm's transform is the same as the spectral curve of the 
monopole defined
in Section \ref{sec:spectral}.

Standard methods from integrable systems can  be used to solve Nahm's
equations using the curve $S$ and some additional structure. Indeed
in \cite{Hit1} Hitchin uses this approach to determine exactly which
spectral curves correspond to mono\-poles.

One of the important properties of Nahm's equations is that it
is straightforward to define a monopole from a solution of Nahm's
equations plus some boundary conditions. Given such a solution
and a point $(x^1, x^2, x^3)$ in $\R^3$ we define
\begin{align*}
T = \sum_{i=1}^3 T_i \sigma^i
\intertext{and}
x = \sum_{i=1}^3 x^i \sigma^i.
\end{align*}
Now define $E_x$ to be the $L^2$ kernel  of
the operator
$$
D_x = \frac{d\phantom{z}}{dz} - T - x.
$$
It can be shown that this is two dimensional.
We define the connection and Higgs field by choosing an orthonormal
basis $(v_1, v_2)$ for $E_x$ and letting
\begin{align*}
\Phi(v_a) = \sum_{b=1}^2 \int_{-1}^1 \left(v_b, z v_a\right) dz
\intertext{and}
A_i(v_a) =  \sum_{b=1}^2 \int_{-1}^1 \left(v_b, \frac{d v_a}{dx^i}\right) dz.
\end{align*}

This $(A, \Phi)$ define a monopole if the $T_i$ satisfy Nahm's
equations with the appropriate boundary conditions.  Moreover if
the solution of Nahm's equations came from a monopole this
construction returns the same monopole.
For more details see \cite{Nah}.

\section{Rational maps}
Although spectral curves are a useful invariant of monopoles it is
very difficult given a spectral curve, for example a polynomial
as in \eqref{eq:spectralcurve}, to determine if it is
the spectral curve of a monopole.  Another approach to
monopoles is the {\em rational map} of a monopole. By a rational
map we mean a meromorphic function $R(z) \colon \C \to \C$
of the form
$$
R(z) = \frac{p(z)}{q(z)}
$$
where $p$ and $q$ are polynomials with no common divisor.
We will be interested first in {\em based} rational maps, that is those
which send $\infty$ to $0$ and have degree $k$. For a degree $k$ based
rational map we  require that $q$ is monic of degree $k$ and that
$p$ has degree strictly less than $k$.  Let $\RR^b_k$ be the set of all
  such maps then we have
\begin{theorem}[Donaldson \cite{Don}]
	If we choose an orthogonal splitting of $\R^3$ as $\C \times \R$ then
	there is a diffeomorphism
	$$
	\M_k \simeq \RR^b_k.
	$$
	\end{theorem}

We shall indicate in a moment how to construct a rational map from a
monopole.  It is important to note that the inverse construction from
rational map to monopole is not  known. All that is
known is that there is a map $\M_k \to \RR^b_k$ which can be shown to
be a diffeomorphism.  This is to be contrasted with the case of
spectral curves where, although we have not discussed it, an inverse
construction is  known and with Nahm's construction
where the  inverse construction is quite straightforward.  It seems that what
we gain by knowing that every rational map is the rational map of some
monopole we lose by not knowing how to get back to a monopole given
a rational map.

To define the rational map we choose first the orthogonal splitting.
This amounts to choosing a direction in $\R^3$ and an orthogonal
plane. Now consider Hitchin's equation along each line parallel to
that direction.    Because of the framing
we have
$$
\lim_{t \to \infty} \Phi(0, 0, t) = \frac{1}{\sqrt{2}}
\left(\begin{matrix} i & 0 \\ 0 & -i
\end{matrix}\right).
$$
We choose a gauge so that
$$
\lim_{t \to \infty} \Phi(x, y, t) = \frac{1}{\sqrt{2}}
\left(\begin{matrix} i & 0 \\ 0 & -i
\end{matrix}\right)
$$
for every $(x, y)$.  For each $z = x + iy$ consider the 
one-dimensional subspace
of $\C^2$ of solutions decaying along the line $(x, y, t)$ as
$t\to \infty$.  Up to scale this line is spanned by a vector
$(a(z), b(z))$. It turns out that $b(z)$ is a
polynomial of degree $k$ with roots $\beta^1, \dots, \beta^k$.
We choose the unique polynomial $p(z)$of degree $k-1$  such that
$p(\beta^i) = a(\beta^i)$ for all $i =1, \dots, k$.  Finally
we define $R(z) = p(z)/b(z)$.  Notice that
as $|x|^2 + |y|^2 \to \infty$ the solution which decays as
$t \to -\infty$ blows up as $t \to \infty$ and hence $a(z) \to 0$ so
that map $R(z)$ is based.  Note also that the spectral lines of the
form $(x, y, t)$ have $b(x+iy) = 0$ so that they correspond to the
poles of the rational map.

While an extremely useful invariant of the monopole as every based
rational corresponds to exactly one monopole there are two significant
problems. One we have already noted: the inverse map from rational 
maps to monopoles is not known.  Another is
that the construction breaks symmetry by writing $\R^3$ as $\C
\times \R$ and hence the construction does not preserve the full group
of  rotations of $\R^3$ but instead the group of rotations around the
axis which determines the splitting $\C \times \R$.

There is an alternative approach to rational maps due to Jarvis
\cite{Jar, Jar1}
which avoids the problem of symmetry breaking.  The rational
map is also simpler to define.  For this we pick a point in $\R^3$
and consider Hitchin's equation along each (oriented) line through that point.
Parametrise the lines through the point by $\zeta$ and the
distance along the line by $t$ then the solution that decays
in the forward direction is given by
$$
s(\zeta, t) =
\begin{pmatrix}
	s^0(\zeta, t)  \\
	s^1(\zeta,t)
\end{pmatrix}.
$$
The Jarvis rational map is then defined by
$$
R(\zeta) = \frac{s^0(\zeta, 0)}{s^1(\zeta, 0)}.
$$
To obtain a gauge invariant correspondence we note that the
rational map is only determined up to the action of $SU(2)$
by fractional linear transformations.  Then Jarvis
proves:
\begin{theorem}[Jarvis \cite{Jar, Jar1}]
If we choose a point in $\R^3$	there is a diffeomorphism
	$$
	\M_k/S^1 \simeq \RR_k/SU(2).
	$$
	\end{theorem}
Here $\RR_k$ is the set of all rational maps of
degree $k$ not just based and the space $\M_k/S^1$ is the
unframed moduli space, that is the moduli space
divided by the action of the circle acting by constant gauge
transformations.

The precise relationship between the Jarvis rational map
and the Donaldson rational map has not yet been determined but it is
believed that by moving the point determining the
Jarvis map out to infinity along a line, and suitably
rescaling the map, that the Donaldson map will be recovered.

\section{Conclusion}
We conclude by considering three significant  omissions and
generalisations.

Firstly the moduli space has a natural hyperkaehler metric. 
The physical interpretation of this is as follows. Consider the 
configuration space ${\mathcal C}$ of all pairs $(A, \Phi)$
satisfying the boundary conditions of a monopole. Inside ${\mathcal C}$ 
we have the space ${\mathcal C}_0$ of solutions of the Bogomolny equations 
and the monopole moduli space is the quotient of ${\mathcal C}_0$ by the 
group of all gauge transformations. 
 A solution of the full Yang-Mills-Higgs equations describes a curve
$(A(t), \Phi(t))$ in ${\mathcal C}$  where the $t$ parameter
is  time.  Monopoles, points on ${\mathcal C}_0$, are static solutions 
of the full Yang-Mills-Higgs equations and the minima of a potential 
energy functional on ${\mathcal C}$. 
If we start with a point $(A, \Phi)$ on ${\mathcal C}_0$ 
and give it a small push  then, by conservation of energy, it should
describe a path staying  close to
${\mathcal C}_0$ and solving the full Yang-Mills-Higgs equations. 
In the limit, as the size of the push becomes zero, 
these paths determine geodesics of the natural
hyperkaehler metric. 

It follows that the geodesics on the 
monopole moduli space of this metric should approximate the motion of slowly moving 
monopoles. 
If we start with a monopole near the edge of the moduli space, 
which can be interpreted as a collection of $k$ particles, and follow
it along a geodesic until it emerges again into the region where 
it can be regarded as $k$ particles we will have described a 
scattering 
process which should be an approximation to true $k$-monopole scattering. 
 This idea is due to Manton \cite{Man1} and in 
\cite{AtiHit} Atiyah and Hitchin describe 
the metric on the moduli space of charge two monopoles and certain of 
its geodesics and interpret the results in terms of two particle
monopole scattering. 

Secondly it is possible to generalise all the present discussion to
other compact, simple Lie groups although the Nahm correspondence
only works simply for the classical Lie groups. See for example,
\cite{Mur, HurMur, HurMur1}. It is also possible
to generalise to the loop group and loop group monopoles or calorons
can be interpreted as instantons on $S^1 \times \R^3$ \cite{GarMur}.

Thirdly the underlying space $\R^3$ can be replaced by hyperbolic
three space and many of the results here carried over. This idea
is due to Atiyah   \cite{Ati} and further developed in \cite{BraAus, 
MurSin} and \cite{MurSin2}.

\end{document}